\documentclass[review]{elsarticle}

\usepackage{lineno,hyperref,amsmath}
\modulolinenumbers[5]

\journal{Journal}

\begin{document}

\begin{frontmatter}

\title{Electronic structure, magnetoexcitons and valley polarized electron gas 
in 2D crystals}

\author[address1]{L. Szulakowska\corref{mycorrespondingauthor}}
\cortext[mycorrespondingauthor]{Corresponding author}
\ead{lszul050@uottawa.ca}

\author[address1,address2]{M. Bieniek}

\author[address1]{P. Hawrylak}

\address[address1]{Department of Physics, University of Ottawa, Ottawa, Canada}
\address[address2]{Department of Theoretical Physics, Wroclaw University of Technology, Wroclaw, Poland}

\begin{abstract}
We describe here recent work on the electronic properties, magnetoexcitons and valley polarised electron gas in 2D crystals. Among 2D crystals, monolayer $MoS_2$ has attracted significant attention as a direct-gap 2D semiconductor analogue of graphene. The crystal structure of monolayer $MoS_2$ breaks inversion symmetry and results in K valley selection rules allowing to address individual valleys optically. Additionally, the band nesting near Q points is responsible for enhancing the optical response of $MoS_2$.We show that at low energies the electronic structure of $MoS_2$ is well approximated by the massive Dirac Fermion model. We focus on the effect of magnetic field on optical properties of $MoS_2$. We discuss the Landau level structure of massive Dirac fermions in the two non-equivalent valleys and resulting valley Zeeman splitting. The effects of electron-electron interaction on the valley Zeeman splitting and on the magneto-exciton spectrum are described. We show the changes in the absorption spectrum as the self-energy, electron-hole exchange and correlation effects are included. Finally, we describe the valley-polarised electron gas in $WS_2$ and its optical signature in finite magnetic fields.
\end{abstract}


\end{frontmatter}

\linenumbers

\section{Introduction}

There is currently interest in the electronic and optical properties of van der Waals (vW) crystals \cite{Yoffe,Besenbacher,Julia,Heinz2010,JuliaNano,Heinz2012,SSComm2012,Geim,XuHeinz,McEuen,Yao,Mak2018,Urbaszek,Zhao,
proximity,Huang,HawrylakNature,Jadczak2017,JadczakSub,PRB2018,Goerbig,Hawrylak1993,SSComm1993,Potemski,CastroNeto,nesting}. Bulk van der Waals crystals are found to be insulators, metals, ferromagnets, superconductors and semiconductors. vW crystals are built of weakly bound atomic planes, hence atomic layers from different vW crystals can now be peeled off and reassembled into new materials with properties not readily available in nature \cite{Geim,Zhao,proximity,Huang}. When bulk vW crystal is reduced to a single atomic layer, the properties can change drastically. For example, bulk $MoS_2$, a well-known transition metal dichalcogenide (TMDC), is an indirect gap semiconductor while a single layer is an example of a truly two-dimensional, direct gap, semiconductor. TMDCs share hexagonal lattice with graphene and the low energy spectra can be understood in terms of massive Dirac Fermions \cite{Zhao}. The two nonequivalent valleys can be addressed optically \cite{Mak2018,Urbaszek,HawrylakNature}, topology leads to valley spin Hall effect \cite{XuHeinz,McEuen} and electron-electron interactions can lead to a broken symmetry valley polarized electronic state \cite{HawrylakNature}. The absorptivity of 2D TMDC layers is very strong due to band nesting \cite{CastroNeto,nesting} and excitonic effects are pronounced due to reduced dimensionality and screening \cite{Zhao}. Here we describe some of our work toward the understanding of the electronic and optical properties of semiconductor TMDCs \cite{XuHeinz,PRB2018,CastroNeto}.

\begin{figure}
    \centering
    \includegraphics[width=0.8\textwidth]{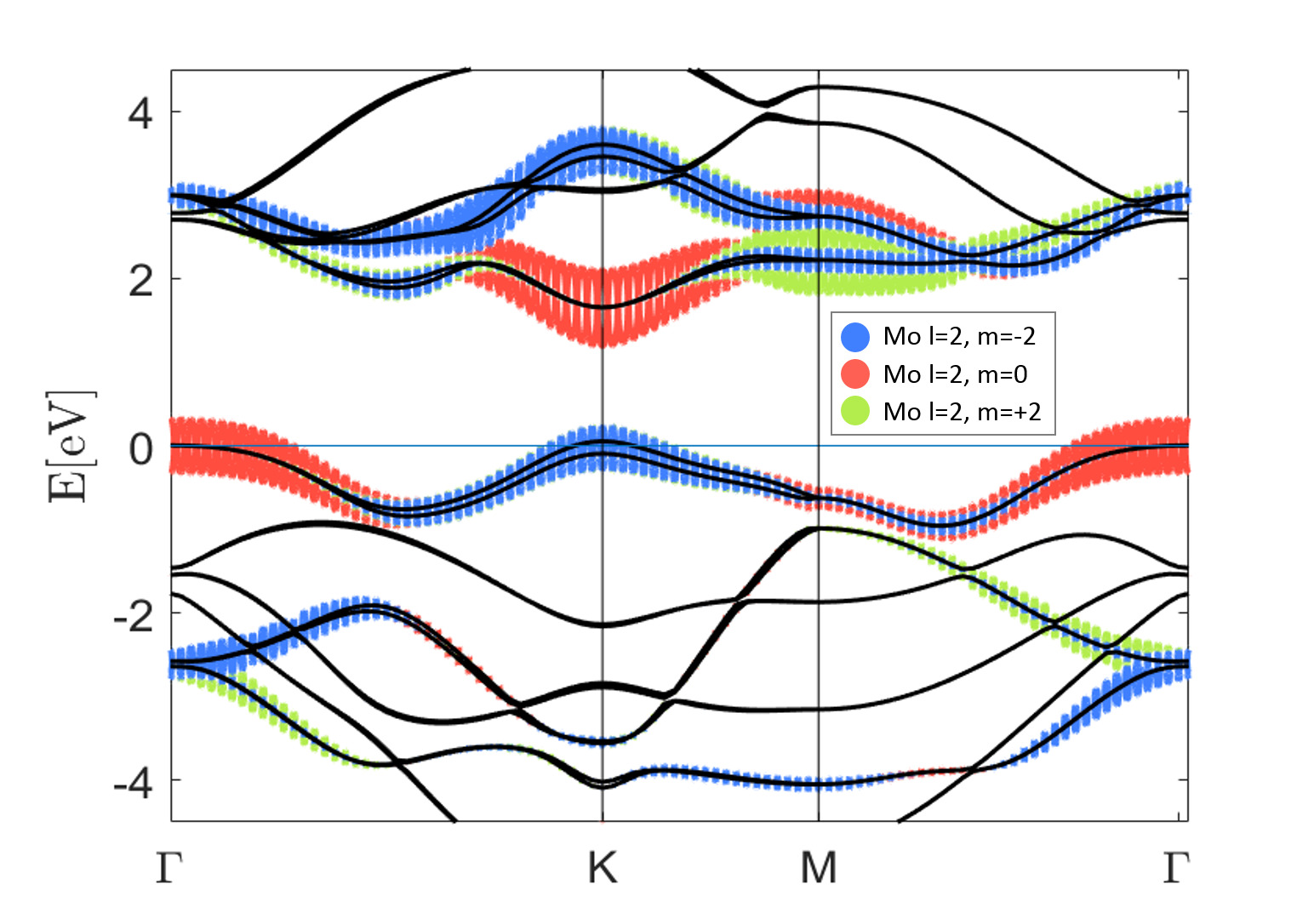}
    \caption{Electronic band structure of a single layer of $MoS_2$ with SO included calculated in DFT. Contributions from d-type orbitals are marked with colors.} \label{orbs}
\end{figure}

\section{Electronic structure of a monolayer of $MoS_2$}

We start with the electronic structure of a best known TMDC, $MoS_2$. Fig. \ref{orbs} shows the ab-initio band structure of a single-layer of $MoS_2$ obtained with the Abinit package \cite{SSComm2012,HawrylakNature,PRB2018}. $MoS_2$ has a layered structure formed by a triangular lattice of Mo atoms sandwiched between planes of triangularly arranged S atoms, resembling honeycomb structure of graphene when viewed from above. Analogous to graphene the first Brillouin zone is hexagonal, with 6 K points at the six corners. Like in graphene, the 6 K points can be divided into two groups of 3 equivalent points., one around K and a second around -K. Because of broken inversion symmetry, K and -K points are not equivalent. Moreover, as seen in Fig. \ref{orbs}, in contrast with graphene, $MoS_2$ is a direct gap semiconductor, with both the conduction and valence band edges located in the K valleys. $MoS_2$ exhibits an indirect-direct gap transition as a function of the number of layers. It is indirect for bulk all the way down to double layer and becomes a direct gap semiconductor only for a single-layer, with direct gap corresponding to optical transitions in the visible range. The Kohn-Sham energy gap in Fig. \ref{orbs} corresponds to $Eg=1.79 eV$. Reduction of $MoS_2$ to a single layer also breaks the inversion symmetry, which gives rise to valley-dependent optical selection rules: transitions in K and -K valleys have been demonstrated to couple to oppositely circularly polarized light \cite{Mak2018,Urbaszek,HawrylakNature}. 

Fig. \ref{orbs} shows the band structure of a single-layer of $MoS_2$ with contributions from different atomic d-orbitals of the metal, Mo, shown in different colours. Top of the valence band $(m_d=\pm2)$ and the bottom of the conduction bands $(m_d=0)$ are built mainly from different Mo d-type orbital, unlike in graphene for which the low energy bands are dominated by $p_z$ orbitals of the C atom. Strong contribution of the d-type orbitals in $MoS_2$ produces large SO-coupling resulting in large spin splitting of the valence band. This splitting, of the order of 150meV at K points, results in two classes of optical transitions, A and B \cite{Heinz2010,JuliaNano,SSComm2012,HawrylakNature,PRB2018}. An important feature of the band structure is the existence of the additional conduction band minima at Q points. Such shape of the conduction band implies that the conduction and valence bands run parallel as a function of k, leading to band nesting, which significantly enhances absorption by TMDCs compared to graphene \cite{PRB2018}. 

\section{Tight-binding model for $MoS_2$ and massive Dirac fermions}

In order to understand important features of the band structure of $MoS_2$ a simple, tight binding (TB), model is needed. Guided by our ab-initio calculations \cite{SSComm2012} we construct such a model \cite{PRB2018} starting with Mo d- orbitals $\varphi_{m_d=\pm2,0}$,  even with respect to the plane and even combination of S p- atomic orbitals  $\varphi_{m_p=±1,0}$ . As in graphene, we construct Bloch wavefunctions for each d orbital of metal sublattice A and each p orbital of sulfur sublattice B:
\begin{align}
\Psi_{A,m_d}\left(k,r\right)=\frac{1}{\sqrt{N_{UC}}}\sum_{i=1}^{N_{UC}}{e^{ikR_{A,i}}\varphi_{m_d}\left(r-R_{A,i}\right)}, \nonumber \\
\Psi_{B,m_p}\left(k,r\right)=\frac{1}{\sqrt{N_{UC}}}\sum_{i=1}^{N_{UC}}{e^{ikR_{B,i}}\varphi_{m_p}\left(r-R_{B,i}\right)}
\end{align}
where $N_{UC}$ is the number of unit cells. We next construct the tunneling matrix elements between two sublattices in analogy to graphene:
\begin{align}
\left\langle A,m_d,k\middle|\hat{H}\middle|\ B,m_p,k\right\rangle&=\int dr\varphi_{m_d}^\ast\left(r\right)V_A\left(r\right)\bigg(e^{ikR_{B1}}\varphi_{m_p}\left(r-R_{B1}\right) \nonumber \\
& +e^{ikR_{B2}}\varphi_{m_p}\left(r-R_{B2}\right)+e^{ikR_{B3}}\varphi_{m_p}\left(r-R_{B3}\right)\bigg),
\end{align}
where $V_A\left(r\right)$ is the potential on sublattice A and $R_{B1},R_{B2},R_{B3}$ are positions of three nearest-neighbors measured from metal atom A. Evaluating this matrix element at point K of the BZ gives matrix element
\begin{equation}
\left\langle A,m_d,K\middle|\hat{H}\middle|\ B,m_p,K\right\rangle=\left(1+e^{i\left(1-m_d+m_p\right)\frac{2\pi}{3}}+e^{i\left(1-m_d+m_p\right)\frac{4\pi}{3}}\right)V_{pd}\left(m_d,m_p\right), \label{mos2mat}
\end{equation}
where$ V_{pd}\left(m_d,m_p\right)$ is a Slater-Koster matrix element for nearest-neighbour Mo-S tunneling. In graphene the same matrix element for tunneling from Pz orbitals of sublattice A to nearest neighbor Pz orbitals of sublattice B reads:
\begin{equation}
\left\langle A,m_p,K\middle|\hat{H}\middle|\ B,m_p,K\right\rangle=\left(1+e^{i\left(1\right)\frac{2\pi}{3}}+e^{i\left(1\right)\frac{4\pi}{3}}\right). \label{graphmat}
\end{equation}

\begin{figure}[h]
    \centering
    \includegraphics[width=0.8\textwidth]{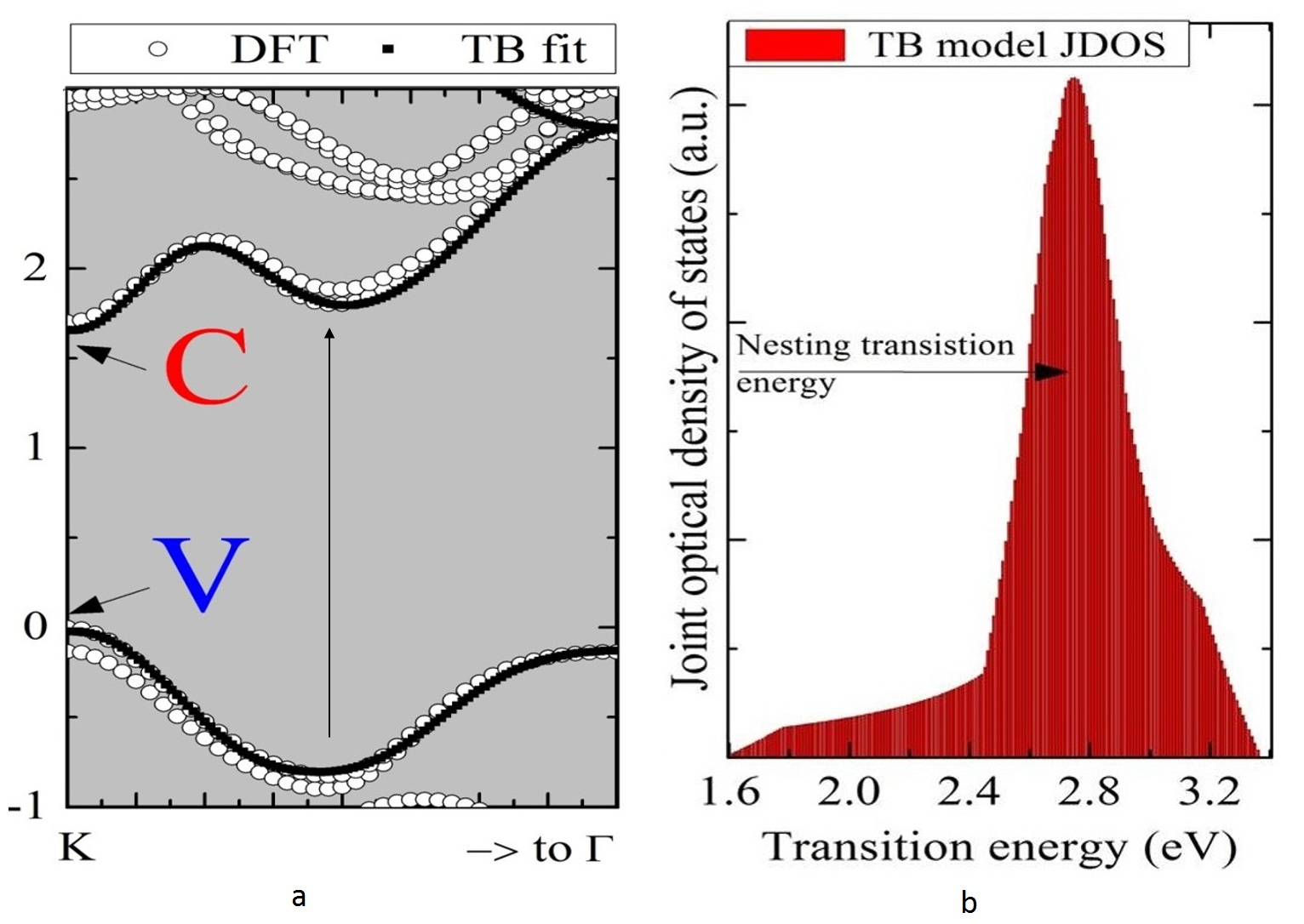}
    \caption{A) Fit of the NNN TB model to DFT band structure. B) Joint optical density of states obtained from the TB model with the peak originating in the band nesting near K and Q points. }\label{jdos}
\end{figure}

The interference between tunneling to three nearest neighbor atoms in graphene at K points, eq. (\ref{graphmat}), leads to vanishing tunneling matrix element and closing of the energy gap at K points of the BZ.  The same process involving Mo d- orbitals and nearest neighbor p orbitals of sulfur in $MoS_2$, eq. (\ref{mos2mat}), involves additional phase factors $\left(m_d-m_p\right)2\pi/3$  because of the different angular momenta of the d-orbitals of Mo and p-orbitals of sulfur dimer $S_2$. In $MoS_2$ the d-orbitals are degenerate in energy and tunneling to nearest neighbor sulfur atoms must remove this degeneracy and hence not close but open the energy gap at K points. Hence we require that the tunneling matrix element, eq. (\ref{mos2mat}), be nonzero at different K points. This is possible only for specific sets of $m_d$ and $m_p$ orbitals for a given K point. For a K-point $K=\frac{2\pi}{a}\left(0,\pm\frac{2}{3}\right)$ finite tunneling matrix element \cite{PRB2018} can only be obtained  for 3 sets of orbitals: $(m_d=0,m_p=-1),(m_d=2,m_p=+1),(m_d=-2,m_p=0)$ resulting in a 6-band TB model. Coupling of each of the d-orbitals to different set of p-orbitals removes the degeneracy of d-orbitals. As seen in Fig. \ref{orbs}. the lowest energy state contributing to the top of the valence band is $m_d=-2$ orbital, the next in energy is the bottom of the conduction band made of the $m_d=0$ orbital and the $m_d=+2$ is high in the conduction band. By contrast, different orbitals are coupled at the  $\Gamma$ point $(m_d=0,m_p=0),(m_d=2,m_p=-1),(m_d=-2,m_p=1)$. As seen in Fig. \ref{orbs}., this different coupling of the two sublattices leads to a different ordering of bands at K and $\Gamma$ points. For example, $m_d=0$ orbital contributes to the bottom of the conduction band at K point but forms the top of the valence band at $\Gamma$ points. Hence crossing of energy levels of different d-orbitals is necessary when moving from K to $\Gamma$ points. This crossing of orbitals results in complex band structure, a set of second minima in the conduction band at Q points and band nesting. We illustrate these effects in Fig. \ref{jdos}. a which shows the valence and conduction energy bands from K to $\Gamma$ points, from the tight binding model with nearest and next nearest neighbor tunneling and ab-initio DFT results. We see the TB model reproducing very well the gap and dispersion at K point as well as the appearance of the second minimum at Q point. The second minimum at Q point due to crossing of different d-orbitals leads to band nesting, i.e., the energy of conduction band parallels the valence band energy dispersion.  The nesting of conduction and valence band produces enhancement in the joint optical density of states shown in Fig. \ref{jdos}b. Importantly, both nearest and next nearest neighbor tunneling processes need to be present in the Hamiltonian to correctly describe the electronic properties of $MoS_2$. For example, without the Mo-S tunneling the system reduces to two decoupled triangular lattices while without the Mo-Mo tunneling we get incorrect position of gaps and masses.

The full six-band TB Hamiltonian can be reduced to a two-band effective mass model Hamiltonian at $k=K+q$  in the basis of the conduction and valence band states at $K\left(\tau=1\right)$ and $-K\left(\tau=-1\right)$:
\begin{equation}
H\left(q\right)=at\left(
\begin{matrix}
0&\tau q_x-iq_y\\\tau q_x+iq_y&0\\
\end{matrix}
\right)+\frac{\Delta}{2}\left(
\begin{matrix}
1&0\\0&-1\\
\end{matrix}
\right). \label{mdfH}
\end{equation}
The Hamiltonian in eq. (\ref{mdfH}) describes massive Dirac fermions (mDfs) \cite{PRB2018,Goerbig} as excitations of the TMDC, with 
$a=3.193\AA,t=1.4677eV$ and $\Delta=1.6848eV$  \cite{PRB2018} extracted from TB and ab-initio calculations.

\section{$MoS_2$ response to external magnetic field}

We now describe Landau quantisation of the Dirac fermion energy levels. The energy spectrum can be obtained by transforming the momentum operator in the massive Dirac Hamiltonian, eq. (\ref{mdfH}), into creation and annihilation operators \cite{Urbaszek,PRB2018,Goerbig,Hawrylak1993}. For K valley the mDF Hamiltonian in B reads:
\begin{equation}
H=\frac{\Delta}{2}\left(
\begin{matrix}
1&0\\0&-1\\
\end{matrix}
\right)+v\left(
\begin{matrix}
0&-i\hat{a}\\i{\hat{a}}^\dag&0\\
\end{matrix}
\right), \label{magn}
\end{equation}
where $v=\frac{\sqrt2v_F}{l_0}$ , $v_F$ is the Fermi velocity $v_F=ta$ and we choose a symmetric gauge $A=\frac{B}{2}\left(-y,x,0\right)$. Diagonalizing the mDF Hamiltonian in eq. (\ref{magn}) we find energy levels for the conduction band ($E>0$) and valence band ($E<0$) as \cite{PRB2018,Goerbig}.
\begin{equation}
E_n^{+/-}=\pm\sqrt{\left(\frac{\Delta}{2}\right)^2+v^2n}
\end{equation}
with corresponding eigenstates $\Psi_{nm}^{+/-}\left(K\right)=\left(
\begin{matrix}
\alpha_n^{+/-}\left.|n-1,m\right\rangle\\\beta_n^{+/-}\left.|n,m\right\rangle\\
\end{matrix}\right)$. Here +/- correspond to positive and negative energy solutions. The eigenvectors are spinors in the basis of conduction and valence band states at K point, consisting of LL states differing by 1. The energy spectrum consists of positive and negative energy levels symmetrically placed in both valleys for $n>0$ and an asymmetric $0^{th}$ LL for $n=0$ placed at the top of the valence band in K valley and at the bottom of the conduction band in the -K valley. This creates a valley Zeeman splitting for electron in the conduction band as
\begin{equation}
\Delta_{Vz}^C=E_1^{CB}\left(K\right)-E_0^{CB}\left(-K\right)\approx\Delta\left(\frac{v_F}{\Delta l_0}\right)^2\hbar\omega_c.
\end{equation}

Even greater asymmetry is apparent if spin-orbit splitting is included. The Hamiltonian for both spin up and spin down in the K valley reads:
\begin{equation}
H_{SO}^K=\left(
\begin{matrix}
\frac{\Delta}{2}-\frac{\Delta_{SO}^C}{2}&-iv\hat{a}&0&0\\iv{\hat{a}}^\dag&-\frac{\Delta}{2}-\frac{\Delta_{SO}^V}{2}&0&0\\0&0&\frac{\Delta}{2}+\frac{\Delta_{SO}^C}{2}&-iv\hat{a}\\0&0&iv{\hat{a}}^\dag&-\frac{\Delta}{2}-\frac{\Delta_{SO}^V}{2}\\
\end{matrix}
\right) \label{Hspin}
\end{equation}
where $\Delta_{SO}^{C/V}$ is the spin splitting for the conduction (valence) band. The solutions of the Hamiltonian in eq. (\ref{Hspin}) are
\begin{equation}
E_n^{+/-}=\sigma\frac{\Delta_{SO}^C+\Delta_{SO}^V}{4}\pm\sqrt{\left(\frac{\Delta+\sigma\frac{\Delta_{SO}^C-\Delta_{SO}^V}{2}}{2}\right)^2+v^2n}.
\end{equation}
If we assume that $\Delta_{SO}^C\ll\Delta_{SO}^V$, both positive and negative energy levels will split due to the mixing between conduction and valence band LLs and due to large $\Delta_{SO}^V$.  Only the $0^{th}$ LL at the bottom of the conduction band in the -K valley will have a negligible spin splitting, enhancing the asymmetry between the valleys.

\section{Optical properties of massive Dirac Fermions}

We now discuss the optical properties of massive Dirac Fermions in $MoS_2$ in a magnetic field \cite{Yao,Jadczak2017,Goerbig,Potemski}. We start with inclusion of e-e interactions into the massive Dirac Fermion model. With index $i$ including all the quantum numbers of the massive Dirac Fermions, including the two nonequivalent valleys, and $c\left(c^\dag\right)$ being the annihilation(creation) operators, the Hamiltonian for interacting massive Dirac Fermions in B reads:
\begin{equation}
{\hat{H}}_{CI}=\sum_{i\sigma}{\varepsilon_{i\sigma}{\hat{c}}_{i\sigma}^\dag{\hat{c}}_{i\sigma}}+\frac{1}{2}\sum_{ijkl\sigma\sigma^\prime}{V_{ijkl}{\hat{c}}_{i\sigma}^\dag{\hat{c}}_{j\sigma^\prime}^\dag{\hat{c}}_{k\sigma^\prime}{\hat{c}}_{l\sigma}}. \label{HCI}
\end{equation}
Here the first term describes the single particle spectrum of mDf and the second term describes their interaction. In eq. (\ref{HCI}) $V_{ijkl}$    are the two-body matrix elements  $V_{ijkl}=\left\langle\Psi_i\Psi_j\middle|\hat{V}\middle|\Psi_k\Psi_l\right\rangle$, evaluated  using mDF wavefunctions $\Psi_i=\Psi_{nm}^{+/-}\ =\alpha_n^{+/-}\left.|n-1,m\right\rangle\left.|C\right\rangle+\beta_n^{+/-}\left.|n,m\right\rangle\left.|V\right\rangle$ and two-body interaction potential $V(r,r')$. Here, for comparison with 2D electron gas, we use bare Coulomb interaction $V(r,r')=\frac{e^2}{|r-r'|}$ Because conduction and valence band wavefunctions are a linear combination of two different Landau levels in the valence $\left.|V\right\rangle$ and conduction $\left.|C\right\rangle$ band, the expression includes many terms characterized with contribution from Landau level envelope and rapidly oscillating conduction and valence wavefunctions at K-point. We illustrate this fact by showing the electron-hole attraction Coulomb matrix element for hole in the $0^{th}$ Landau level $\beta_0^- \left.|0,m\right\rangle$ and electron in first excited level $\alpha_1^+\left.\ |0,m\right\rangle\left.\ |C\right\rangle+{\beta_1^+}^\ast\left.\ |1,m\right\rangle\left.\ |V\right\rangle$ :
\begin{align}  \label{eh}
\left\langle\Psi_{1,m}^+\Psi_{0,m}^-\middle|\hat{V}\middle|\Psi_{0,m}^-\Psi_{1,m}^+\right\rangle&=\left({\alpha_1^+}^\ast\left\langle0,m|\right.\left\langle C|\right.+{\beta_1^+}^\ast\left\langle1,m|\right.\left\langle V|\right.\right)\cdot\\ \nonumber
&\left({\beta_0^-}^\ast\left\langle0,m|\right.\left\langle V|\right.\right)\hat{V}\left(\beta_0^-\left.\ |0,m\right\rangle\left.\ |V\right\rangle\right)\cdot\\ \nonumber
&\left(\alpha_1^+\left.\ |0,m\right\rangle\left.\ |C\right\rangle+{\beta_1^+}^\ast\left.\ |1,m\right\rangle\left.\ |V\right\rangle\right)\\ \nonumber
&=\left|\alpha_1^+\right|^2\left|\beta_0^-\right|^2\left\langle0,m;0,m\ \middle|\left\langle C\ V\middle|\hat{V}\middle|\ V\ C\right\rangle\middle|0,m;0,m\right\rangle\\ \nonumber
&+\left|\beta_1^+\right|^2\left|\beta_0^-\right|^2\left\langle1,m;0,m\ \middle|\left\langle C\ V\middle|\hat{V}\middle|\ V\ C\right\rangle\middle|0,m;1,m\right\rangle.\\ \nonumber 
\end{align}
We see that Coulomb matrix element is a product of band contribution $\left\langle C V\middle|\hat{V}\middle| V C\right\rangle$ and envelope function contributions. Because of massive Dirac fermion nature of our quasi-electron, there are contributions from 0 and 1 Landau levels. The value of $V_{CVVC}$ in eq. (\ref{eh}) is the strength of the interaction between electrons in conduction and valence bands and $\left\langle0,m;0,m\ \middle|\hat{V}\middle|0,m;0,m\right\rangle$ are Coulomb matrix elements for 2D electrons in a magnetic field. \cite{Hawrylak1993}

\begin{figure}[h]
    \centering
    \includegraphics[width=0.8\textwidth]{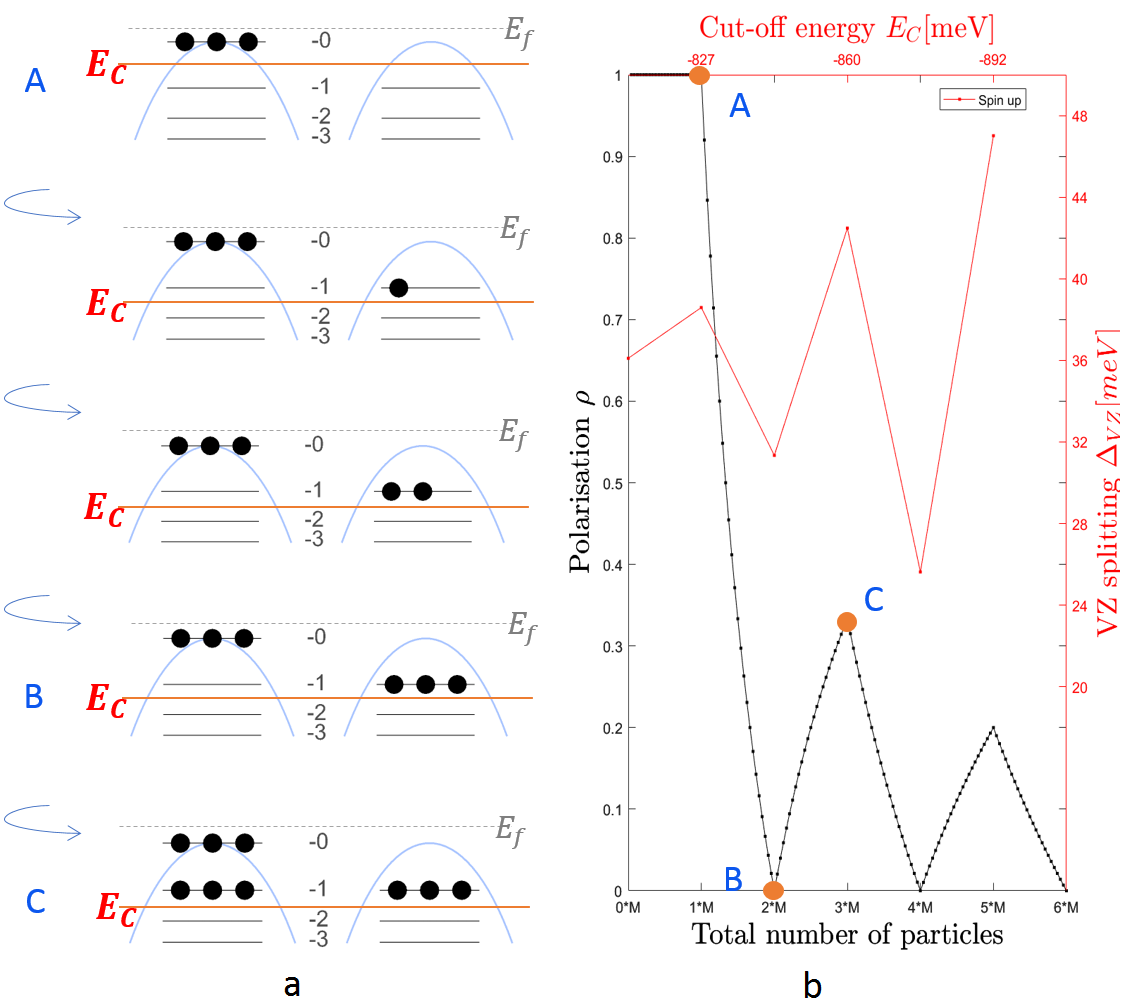}
    \caption{ A) Demonstration of the valley polarization in mDF electron gas. Setting the cut-off energy together with the number of particles created polarized or depolarised GS. B) Valley Zeeman splitting vs. the valley polarization. }\label{valleypol}
\end{figure}

To describe the ground state of weakly interacting mDf we populate the valence band of mDf LLs with electrons in both K and –K valleys to form a Hartree-Fock groundstate (HFGS) $\left.\ |GS\right\rangle=\prod_{\lambda<\lambda_F}{\left(c_{\lambda\left(-K\right)}^V\right)^\dag\left.|0\right\rangle}\prod_{\lambda<\lambda_F}{\left(c_{\lambda\left(K\right)}^V\right)^\dag\left.|0\right\rangle}$ where $\lambda$ corresponds to a collective index $\lambda=\left(n,m,\sigma\right),$  excluding the valley index. In the presence of valley Zeeman splitting, increasing the total number of particles and hence energy cut-off $E_C$ changes the number of particles in each valley, as shown in Fig.  \ref{valleypol}a. Due to asymmetry in the structure of LLs some values of $E_C$ correspond to unequal number of particles in both valleys and the system develops valley-polarisation  \cite{Mak2018}: $\rho=\frac{N_K-N_{-K}}{N}$, where $N_{K\left(-K\right)}$ is the number of particles in valley K (-K) and N is the total number of particles in both valleys. The oscillation of the valley polarization is shown in Fig. \ref{valleypol}b.

Once we have the groundstate we form a single excitation in a given valley, an electron-hole pair, from the GS of the form:
\begin{equation}
\left.|ij\right\rangle=\left(c_j^C\right)^\dag c_i^V \left.|GS\right\rangle.
\end{equation}
Such a pair is not the eigenstate of the interacting Hamiltonian. We next form a magneto-exciton as a linear combination of excited pairs. $\left.|\mathrm{\Phi}_f\right\rangle=\sum_{ij}{A_{ij}^f\left(c_j^C\right)^+c_i^V\left.|GS\right\rangle}$, The exciton wavefunction is obtained by solving the Bethe –Salpeter equation for amplitudes $A_{ij}$:
\begin{equation}
\left(\left(\varepsilon_j+\Sigma_j\right)-\left(\varepsilon_i+\Sigma_i\right)\right)A_{ij}+\sum_{kl}{\left(V_{ilkj}-V_{iljk}\right)A_{kl}}=EA_{ij}
\end{equation}
where $\Sigma_i$ is the exchange self-energy of mDf in the valence band and $\Sigma_j$ is the exchange self-energy of electron in the conduction band due to filled valence band:
\begin{align}
\Sigma_j&=-\sum_{\lambda<\lambda_F}{\left\langle\left.\ \mathrm{\Psi}_j^+\right|\right.\left\langle\left.\ \mathrm{\Psi}_\lambda^-\right|\right.\hat{V}\left.\ \left|\mathrm{\Psi}_j^+\right.\right\rangle\left.\ \left|\mathrm{\Psi}_\lambda^-\right.\right\rangle},\quad \\\nonumber
\Sigma_i&=-\sum_{\lambda<\lambda_F}{\left\langle\left.\ \mathrm{\Psi}_i^-\right|\right.\left\langle\left.\ \mathrm{\Psi}_\lambda^-\right|\right.\hat{V}\left.\ \left|\mathrm{\Psi}_i^-\right.\right\rangle\left.\ \left|\mathrm{\Psi}_\lambda^-\right.\right\rangle},
\end{align}
Once the exciton wavefunction $\left.|\mathrm{\Phi}_f\right\rangle$  and energy levels $E_f$ are obtained from the BSE, the absorption spectrum is obtained from the Fermi’s golden rule:
\begin{equation}
A\left(\omega\right)=\sum_{f}\left|\left\langle\mathrm{\Phi}_f\ \middle|{\hat{P}}^+\middle|\ G\ S\right\rangle\right|^2\delta\left(\hbar\omega-\left(E_f-E_{GS}\right)\right).
\end{equation}
Here ${\hat{P}}^+=\sum_{ij}{d_{ij}\left(c_j^C\right)^+c_i^V}$ is the interband polarization operator corresponding to photon absorption, $d_{ij}=\alpha_n^{C^\ast}\beta_{n^\prime}^V$ is the dipole moment and the final state is$ \left.|\mathrm{\Phi}_f\right\rangle=\sum_{ij}{A_{ij}^f\left(c_j^C\right)^+c_i^V\left.|GS\right\rangle}$.  The dipole moments satisfy the following selection rules  $d_{ij}: \Delta \sigma=0,\Delta m=0$ and $\Delta n=\pm1$.

\section{Magneto-excitons of massive Dirac Fermions. }

We now discuss massive Dirac Fermion magneto-exciton spectra for a single valley. We start by determining self-energy. To calculate self- energy we need to populate Landau levels in the valence band of both valleys. But only +K valley contains the $n=0$ Landau level hence there is Valley Zeeman splitting in the valence band. Fig. \ref{valleypol} shows the population of LL levels for noninteracting and interacting mDf.

The self-energy renormalizes the LLs in conduction and valence band which affects the valley Zeeman splitting. It shows oscillatory behavior, following the valley polarization as shown in Fig. \ref{valleypol}b. For an unpolarized case it decreases with the number of particles N and for polarized case it increases in value.

\begin{figure}[h]
    \centering
    \includegraphics[width=1.0\textwidth]{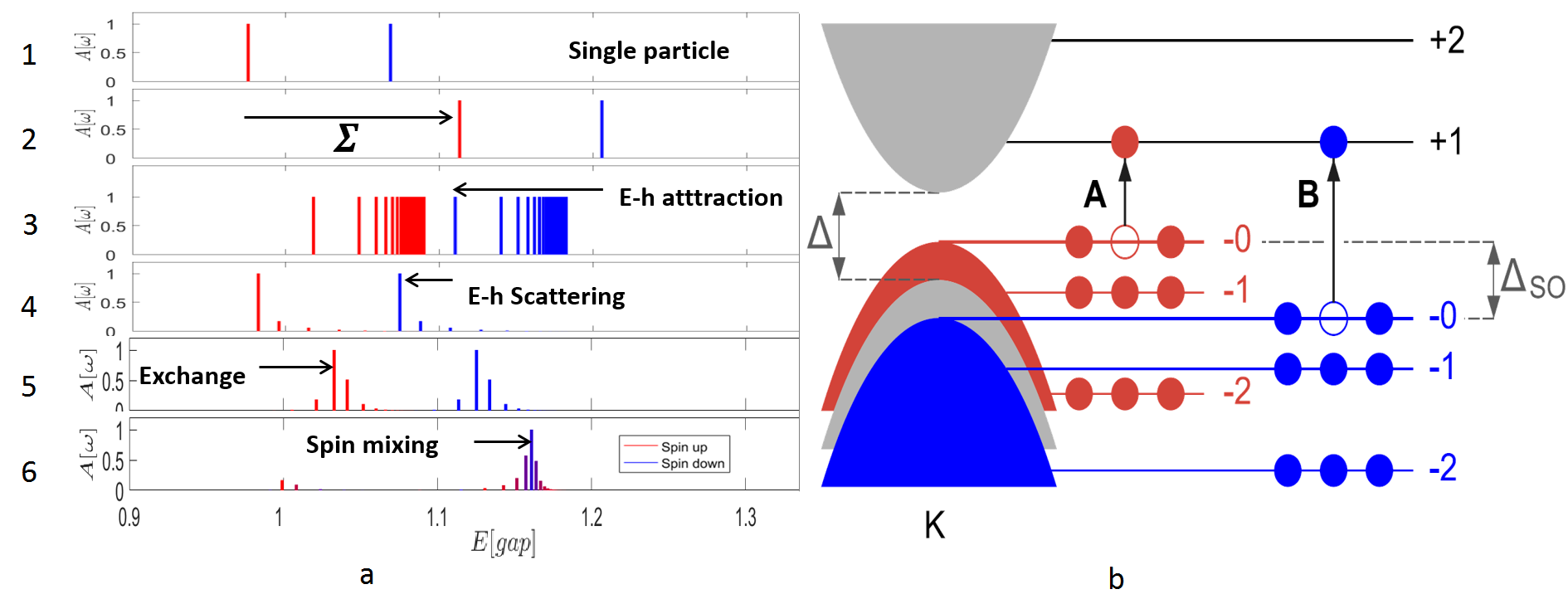}
    \caption{ A) Absorption spectra for different contributions in the Hamiltonian. Colors stand for spins. B) Diagram of the A and B exciton configurations. Colors have the same meanings on both graphs.} \label{magneto}
\end{figure}

We now discuss how these electronic properties can be detected in an optical measurement. After populating the valence band LLs we compute a single exciton (Fig. \ref{magneto}) by solving the BSE.

In Fig. \ref{magneto}a we show the different approximations to the computed absorption spectrum for an example of the GS made of particles in the topmost LL of the valence band.  We start from absorption by noninteracting electrons, Fig. \ref{magneto}a-1. All transitions from filled m states are at the same energy, $E=+\sqrt{\left(\frac{\Delta}{2}\right)^2+v^2}+\frac{\Delta}{2}$ equal to the single particle gap plus contribution from 1st Landau level in the conduction band. Next, we renormalize initial and final states by the the self-energy, Fig. \ref{magneto}a-2. which leads to the blue shift of transition energy. In next step, Fig. \ref{magneto}a-3., we include electron-hole attraction, which leads to a red shift balancing the exchange self-energy contribution and a spread in transition energies. Finally, we allow for correlation effects, i.e., scattering of different electron-hole pair configurations, responsible for renormalization of oscillator strengths of different transitions., Fig. \ref{magneto}a-4. The absorption peaks evolve further as the exchange interaction is switched on and as the spin configurations are allowed to couple (Fig. \ref{magneto}a-5-6). We find the final absorption maximum to be blue-shifted with respect to the single-particle gap. Further studies will include screening of Coulomb interactions and intervalley scattering effects.

\section{Valley-polarised electron gas in $MoS_2$}

\begin{figure}[h]
    \centering
    \includegraphics[width=0.9\textwidth]{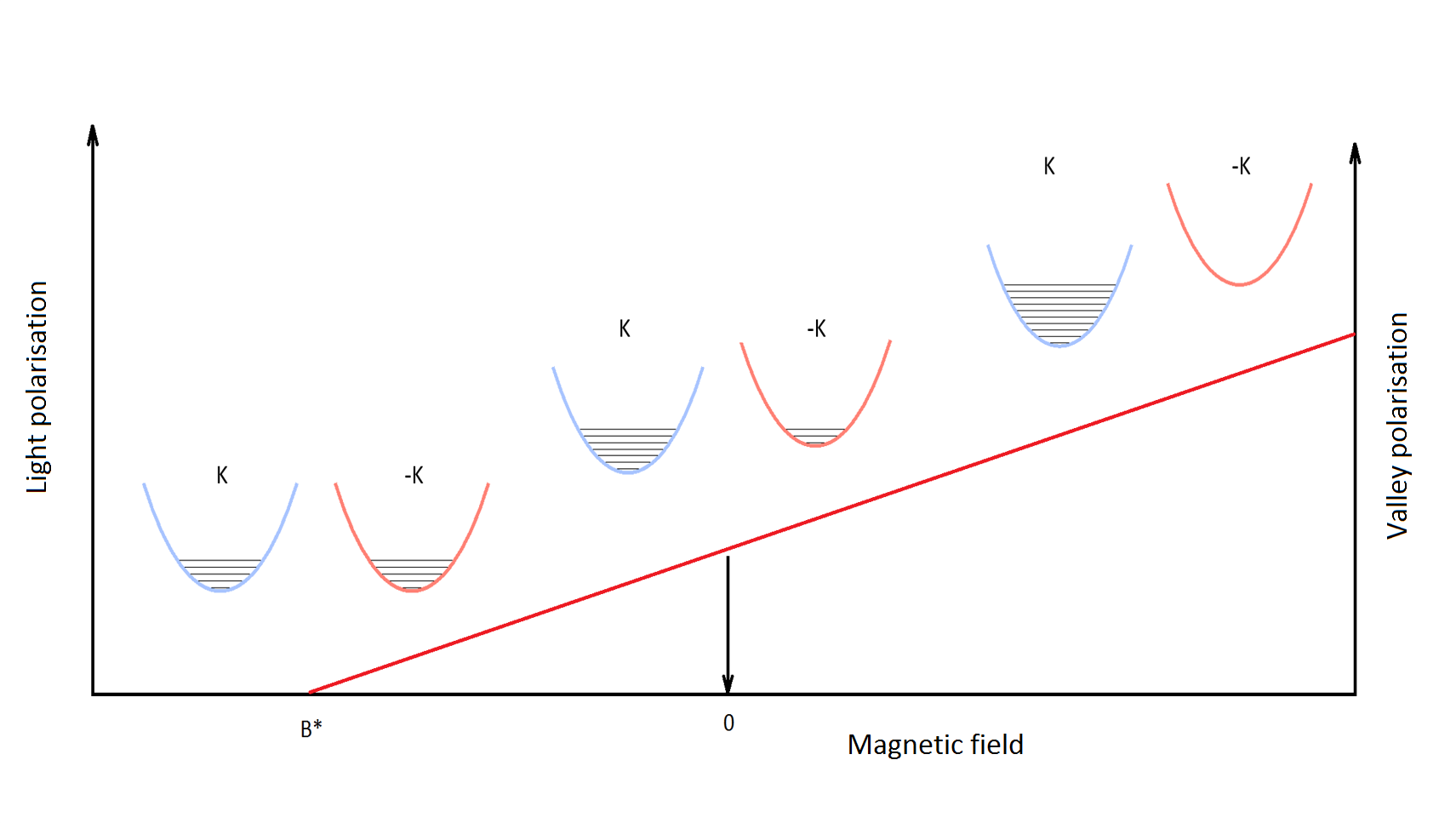}
    \caption{. Valley and light polarisation dependent on magnetic field. For B=0 the valley polarization is non-zero. It vanishes for finite negative magnetic field B. Positive magnetic fields increase the valley polrisation.} \label{final}
\end{figure}

One of the most important properties of TMDCs is the simultaneous presence of two nonequivalent valleys and strong spin orbit coupling resulting in spin-valley locking. Effectively, spin down electrons are found in valley K and spin up electrons in valley -K. Hence we can think of valley up and valley down in the same fashion as we think of spin up or spin down electrons. In the presence of finite electron density we can put half of electrons into valley +K and half into valley -K. In Hartree-Fock approximation the total energy of valley unpolarised state as a function of interparticle separation $r_s$ is given by $E_{tot}(r_s,0)=\frac{1}{r_s}-\frac{8\sqrt{2}}{3\pi}\frac{1}{r_s}$. Alternatively, we can put all electrons into only one, for example, +K, valley with total energy of valley polarized state $E_{tot}(r_s,1)=\frac{2}{r_s}-\frac{16}{3\pi}\frac{1}{r_s}$. We see that we have to pay a penalty in kinetic energy but gain exchange energy in valley polarized state.  Valley polarized electron gas (VPEG) becomes a lower energy state for $r_s>r_s^*=\frac{8(2-\sqrt{2})}{3\pi}$  . The optical transitions in TMDCs are valley selective, e.g., optical recombination from electron in valley +K results in sigma+ polarized photon emission. Hence, even for unpolarised exciting light the emitted light should be circularly polarized. This is what was observed by Scrace et al. \cite{HawrylakNature}. as summarised in Fig. \ref{final}.

At zero external magnetic field the emitted light is circularly polarized. The degree of light and hence, electronic polarization, increases with increasing magnetic field for one direction of B due to spin-valey locking. When direction of B is reversed, the splitting between two valleys is reduced and so is the light polarization. Much theoretical and experimental work is needed to develop a complete understanding of valley polarized electron gas.

Acknowledgment : L.S., M.B. and P.H. acknowledge support from NSERC and uOttawa Research Chair. M.B. acknowledges financial support from National Science Center (NCN), Poland, grant Maestro No. 2014/14/A/ST3/00654.

\section*{References}

\bibliographystyle{plain}
\bibliography{mybibfile}

\begin{thebibliography}{10}
\expandafter\ifx\csname url\endcsname\relax
  \def\url#1{\texttt{#1}}\fi
\expandafter\ifx\csname urlprefix\endcsname\relax\def\urlprefix{URL }\fi
\expandafter\ifx\csname href\endcsname\relax
  \def\href#1#2{#2} \def\path#1{#1}\fi

\bibitem{Yoffe}
G.~Connell, J.Wilson, A.Yoffe, J. Phys. Chem. Solids 30 (1969) 287--296.

\bibitem{Besenbacher}
S.~Helveg, J.~Lauritsen, E.~Lægsgaard, I.~Stensgaard, J.~K. Nørskov, B.~S.
  Clausen, H.~Topsøe, F.~Besenbacher, Phys. Rev. Lett. 84~(5) (2000) 951--954.

\bibitem{Julia}
T.~Li, G.~Galli, J. Phys. Chem. C 111~(1271) (2007) 16192--16196.

\bibitem{Heinz2010}
K.~F. Mak, C.~Lee, J.~Hone, J.~Shan, T.~F. Heinz, Phys. Rev. Lett. 105~(136805)
  (2010) 1--5.

\bibitem{JuliaNano}
A.~Splendiani, L.~Sun, Y.~Zhang, T.~Li, J.~Kim, C.-Y. Chim, G.~Galli, F.~Wang,
  Nano Lett. 10~(4) (2010) 1271--1275.

\bibitem{Heinz2012}
K.~F. Mak, C.~Lee, J.~Hone, J.~Shan, T.~F. Heinz, Nature Nanotechnology 7
  (2012) 494--498.

\bibitem{SSComm2012}
E.~Kadantsev, P.~Hawrylak, Solid State Commun. 152 (2012) 909--913.

\bibitem{Geim}
A.~K. Geim, I.~V. Grigorieva, Nature (London) 499 (2013) 419--425.

\bibitem{XuHeinz}
X.~Xu, W.~Yao, D.~Xiao, T.~Heinz, Nature Physics 10 (2014) 343--350.

\bibitem{McEuen}
K.~F. Mak, K.~L. McGill, J.~Park, P.~L. McEuen, Science 344~(6191) (2014)
  1489--1492.

\bibitem{Yao}
H.~Yu, X.~Cui, X.~Xu, W.~Yao, National Science Review 2 (2015) 57--70.

\bibitem{Mak2018}
K.~F. Mak, D.~Xiao, J.~Shan, Nature Photonics 12 (2018) 451--460.

\bibitem{Urbaszek}
G.~Wang, A.~Chernikov, M.~Glazov, T.~F. Heinz, X.~Marie, T.~Amand, B.~Urbaszek,
  Rev. Mod. Phys. 90~(21001) (2018) 1--25.

\bibitem{Zhao}
C.~Zhao, et~al., Nature Nano. 12 (2017) 757--763.

\bibitem{proximity}
I.~Zutic, A.~Matos-Abiague, B.~Scharf, H.~Dery, K.~Belashchenko, Mater. Today
  12 (2018) 1--23.

\bibitem{Huang}
B.~Huang, et~al., Nature 546 (2017) 270--273.

\bibitem{HawrylakNature}
T.~Scrace, Y.~Tsai, B.~Barman, L.~Schweidenback, A.~Petrou, G.~Kioseoglou,
  I.~Ozﬁdan, M.~Korkusinski, P.~Hawrylak, Nature Nanotechnology 10 (2015)
  603--607.

\bibitem{Jadczak2017}
J.~Jadczak, A.~Delgado, L.~Bryja, Y.~Huang, P.~Hawrylak, Phys. Rev. B
  95~(195427) (2017) 1--9.

\bibitem{JadczakSub}
J.~Jadczak, L.~Bryja, J.~Kutrowska-Girzycka, P.~Kapuściński, M.~Bieniek,
  Y.~S. Huang, P.~Hawrylak, Nature Comm. (submitted 2018).

\bibitem{PRB2018}
M.~Bieniek, M.~Korkusinski, L.~Szulakowska, P.~Potasz, I.~Ozfidan, P.~Hawrylak,
  Phys. Rev. B 97~(085153) (2018) 1--9.

\bibitem{Goerbig}
F.~Rose, M.~O. Goerbig, F.~Piechon, Phys. Rev. B 88~(125438) (2013) 1--7.

\bibitem{Hawrylak1993}
P.~Hawrylak, Phys. Rev. Lett. 71~(3347) (1993) 485--488.

\bibitem{SSComm1993}
P.~Hawrylak, Solid State Comm. 88 (1993) 475--479.

\bibitem{Potemski}
P.~Hawrylak, M.~Potemski, Phys. Rev. B 56~(12386) (1997) 1--9.

\bibitem{CastroNeto}
A.~Carvalho, R.~Ribeiro, A.~C. Neto, Phys. Rev. B 88~(115205) (2013) 1--6.

\bibitem{nesting}
D.~Kozewa, et~al., Nature Communications 5~(4543) (2014) 1--7.

\end{thebibliography}

\end{document}